\documentclass[useAMS,usenatbib]{mn2e}
\usepackage{graphicx,amssymb,color}
\usepackage[normalem]{ulem}

\newcommand{\rev}{ }

\title[Surviving radio-loud white dwarf planets]
{Survivability of radio-loud planetary cores orbiting white dwarfs}

\author[Veras \& Wolszczan]{Dimitri Veras$^{1,2}$\thanks{E-mail: d.veras@warwick.ac.uk}\thanks{STFC Ernest Rutherford Fellow},
Alexander Wolszczan$^{3,4}$
\\
$^{1}$Centre for Exoplanets and Habitability, University of Warwick, Coventry CV4 7AL, UK
\\
$^{2}$Department of Physics, University of Warwick, Coventry CV4 7AL, UK
\\
$^{3}$Center for Exoplanets and Habitable Worlds, Pennsylvania State University, 525 Davey Laboratory, University Park, PA, 16802, USA
\\
$^{4}$Department of Astronomy and Astrophysics, Pennsylvania State University, 525 Davey Laboratory, University Park, PA, 16802, USA
}

\pubyear{2019}

\begin{document}
\label{firstpage}
\pagerange{\pageref{firstpage}--\pageref{lastpage}}
\maketitle

\begin{abstract}
The discovery of the intact metallic planetary core fragment orbiting the white dwarf SDSS J1228+1040 within one Solar radius highlights the possibility of detecting larger, unfragmented conducting cores around magnetic white dwarfs through radio emission. Previous models of this decades-old idea focussed on determining survivability of the cores based on their inward Lorentz drift towards the star. However, gravitational tides may represent an equal or dominant force. Here, we couple both effects by assuming a Maxwell rheological model and performing simulations over the entire range of observable white dwarf magnetic field strengths ($10^3$ -- $10^9$ G) and their potential atmospheric electrical conductivities ($10^{-1}$ -- $10^4$ S/m) in order to more accurately constrain survivability lifetimes. This force coupling allows us to better pinpoint the physical and orbital parameters which allow planetary cores to survive for over a Gyr, maximizing the possibility that they can be detected.  The most robust survivors showcase high dynamic viscosities ($\gtrsim 10^{24}$ Pa$\cdot$s) and orbit within kG-level magnetic fields.
\end{abstract}


\section{Introduction}

Over two decades ago, {\rev \cite{feretal1997} detected striking emission lines in the spectrum of the magnetic white dwarf GD 356. Prompted by this finding,} \cite{lietal1998} proposed that the electromagnetic interactions between a magnetic white dwarf and a conducting planetary core could create a unipolar inductor circuit {\rev similar to Jupiter and Io's \citep{gollyn1969}} that both drags the planet towards the star (a phenomenon known as ``Lorentz drift'') and generates detectable emission. Years later, \cite{wilwu2004} and \cite{wilwu2005} followed up and investigated this idea further by providing additional details on the evolution and radio detection of these planets, based on a model from \cite{wuetal2002}.

Although such planets have not yet been discovered, this possibility is becoming increasingly plausible due to the rapid increase in detections of various features of white dwarf planetary systems. Since 2005, both the available data and our understanding of these systems has matured and grown substantially, as we explain in the next two subsections. Hence, the context in which researchers investigate white dwarf planetary systems has changed.

Further, during this period, investigations of detectable emission from electromagnetic interactions of planets orbiting other types of stars have been numerous; some examples since the year 2015 are \cite{videtal2015}, \cite{fujetal2016}, \cite{katetal2016}, \cite{nicmil2016}, \cite{stretal2017}, \cite{webetal2017}, \cite{dalste2018}, \cite{lynetal2018}, \cite{ogoetal2018}, \cite{strugarek2018}, and \cite{wanloe2019}. Also, these studies incorporated stellar winds, mass loss and/or planetary atmospheres, none of which exist in the scenario that we are considering.

\subsection{Current state of white dwarf planetary systems}

Observable features in white dwarf systems result from substellar objects which avoid engulfment after their host star has left the main sequence \citep{villiv2009,kunetal2011,musvil2012,adablo2013,norspi2013,valras2014,viletal2014,madetal2016,staetal2016,galetal2017,raoetal2018} and then are gravitationally perturbed close to the resulting white dwarf \citep{veras2016a}. After entering the white dwarf's disruption, or Roche radius, an object will break up and form a series of rings or annuli \citep{graetal1990,jura2003,debetal2012,beasok2013,veretal2014c,broetal2017,veretal2017b} usually before accreting on to the white dwarf atmosphere with a particular size distribution \citep{wyaetal2014,veretal2015b}.

These substellar objects which are perturbed towards the white dwarf comprise {\it planets} \citep{debsig2002,veretal2013,voyetal2013,musetal2014,vergae2015,veretal2016,veretal2017a,steetal2018,veretal2018}, {\it asteroids} \citep{bonetal2011,debetal2012,frehan2014,bonver2015,hampor2016,petmun2017,steetal2017,musetal2018,smaetal2018}, {\it comets} \citep{alcetal1986,veretal2014a,stoetal2015,caihey2017} and {\it moons} \citep{payetal2016,payetal2017}. In our solar system, at least five of the major planets will survive until and during the solar white dwarf phase \citep{sacetal1993,dunlis1998,rybden2001,schcon2008,veras2016b}. The inner solar system will likely to be strewn with asteroid belt debris \citep{veretal2014b} while the minor planets in the outer solar system will be re-arranged \citep{veretal2015a,veretal2019a}.

Observations of the consequences of these dynamical interactions are abundant. Accreted planetary debris in white dwarf photospheres are detectable spectroscopically because these stars are dense enough to stratify matter by atomic weight \citep{schatzman1958}; any observed metals heavier than helium then arise from planetary debris, with few exceptions \citep{faretal2010,veretal2019c}. Between one-quarter and one-half of all Milky Way white dwarfs are now thought to contain planetary debris \citep{zucetal2003,zucetal2010,koeetal2014}, a fractional range which is similar to that of the planet occurrence range around main sequence stars \citep{casetal2012}. The composition of the debris provides exclusive and detailed insight into the bulk chemical composition of exo-planetary bodies, even enabling one to link an individual white dwarf with an individual meteorite family in the solar system \citep{zucetal2007,kleetal2010, kleetal2011,ganetal2012,juryou2014,wiletal2015,xuetal2017,haretal2018,holetal2018}.

These metals accrete onto the white dwarf through the aforementioned discs. Their formation has not only been theorized and modelled, but also observed: in fact about 50 such discs have been detected \citep{zucbec1987,farihi2016}. Many more discs are missed simply due to misalignment between our line-of-sight and the plane of the disc \citep{beretal2014,bonetal2017}.  The discs range in radius from about $0.6 R_{\odot}$ to $1.2 R_{\odot}$ and showcase variability \citep{ganetal2008,wiletal2014,xujur2014,faretal2018a,xuetal2018,swaetal2019} and secular changes in spectroscopic features \citep{manetal2016,cauetal2018,denetal2018}, both of which are indicative of dynamical activity. 

All of the discs contain dust and about eight of these discs contain detectable gas \citep{ganetal2006,ganetal2007,ganetal2008,gansicke2011,faretal2012,meletal2012,wiletal2014,guoetal2015}, which is generated from a combination of sublimation and collisions \citep{jura2008,metetal2012,kenbro2017}. The sublimation radius decreases with white dwarf cooling age, and at some point will cross the disruption radius.

In addition to metal pollution and debris discs, minor planets have been observed orbiting white dwarfs. The disintegrating asteroid orbiting WD 1145+017 \citep{vanetal2015} resides a distance of just one Solar radius and has spawned over 20 additional papers. The intact and metallic core object orbiting SDSS J1228+1040 instead resides at a distance of just $0.73R_{\odot}$ and hence cannot be a rubble pile but rather an object of significant density and internal strength \citep{manetal2019}. These minor planets close to white dwarfs are most likely first-generation leftovers from the main sequence phase because second-generation formation in white dwarf debris discs requires massive discs \citep{vanetal2018} and/or a stellar companion \citep{perets2011,schdre2014,voletal2014,hogetal2018}.

\subsection{Planet -- white dwarf interactions}

Overall, the robust state of the observations and theories outlined above has prompted researchers to revisit the potential electromagnetic interactions between a magnetic white dwarf and orbiting bodies \citep{faretal2017,faretal2018b,rapetal2018,broken2019}, especially now that a ferrous core fragment has been discovered orbiting SDSS J1228+1040 \citep{manetal2019}. \cite{broken2019} in particular {\rev have} considered both the Lorentz drift and Ohmic heating of asteroid-sized bodies (including dwarf planets) orbiting a variety of stars, including white dwarfs. The form of the secular orbital decay rate due to Lorentz drift from their Eq. (18) is similar to those reported in \cite{lietal1998}, \cite{wilwu2004} and \cite{wilwu2005}. {\rev Further,} \cite{broken2019} {\rev investigated} in depth the thermal evolution of the asteroids {\rev due to Ohmic heating}. Nevertheless, all four of these studies assume that the Lorentz drift dominates the motion of these planets close to the white dwarf. For cores which are the sizes of terrestrial planets, this assumption is not necessarily justified \citep{bouceb2015,stretal2017}.

\cite{veretal2019b} demonstrated that gravitational tides between a solid planet and a white dwarf play a pivotal role in determining whether a planet migrates into the Roche radius, and if not, in what direction and to what extent does the migration occur. They adopted an arbitrary frequency dependence on the quality dissipation functions and illustrated that the resulting motion may be non-monotonic, and vary the orbit's semimajor axis, eccentricity and inclination.  The non-monotonicity arises from the chaotic interplay between the nonlinear orbital and spin equations of motion \citep{bouefr2019} within a Maxwell rheological model \citep{efroimsky2015}.

\subsection{Plan for paper}

The goal of this paper is to model the interplay between gravitational tides and Lorentz drag in compact white dwarf planetary systems {\rev in} order to more accurately determine survivability timescales. We are keen to determine the planet's survivability as a function primarily of the star's magnetic field strength and atmospheric conductivity, the planet's dynamic viscosity, and the distance from the core to the white dwarf's Roche radius.  We consider only cores large enough to not be destroyed through melting, with the intention that surviving cores can represent targets for radio emission searches \citep{heszar2011,zaretal2018}.

In Section 2 we briefly summarize the relevant theoretical considerations. We report on our simulation outputs in Section 3 before discussing the applicability of our results to other types of {\rev planetary systems and white dwarfs in Sections 4-5. We summarize our results in Section 6}.

\section{Drivers of motion}

In this section we outline the different drivers of motion for a solid metallic or iron-rich object at least the size of Ceres (which we will henceforth refer to as ``planet'') orbiting a white dwarf. 

\subsection{Gravitational tides}

First consider an unmagnetized white dwarf. As the planet approaches the white dwarf from a scattering event at a distance of at least a few au, gravitational tides between the star and planet determine the latter's subsequent motion. The density, spin and internal strength of the planet determines the location of the star's Roche radius, within which that planet will be destroyed. Gravitational tides may push the planet into the Roche radius or push the planet outward, depending on how the orbit is modified.

If the rheologies of both the star and planet are treated within a Maxwell model, then the character of the orbital change is dictated by 14 parameters (see Section 4 of \citealt*{veretal2019b}). These include the four orbital parameters $a$, $e$, $i$, and $i'$ --- which represent the semimajor axis, eccentricity, inclination with respect to the planet's equator, and inclination with respect to the star's equator --- and the following five physical parameters for both the planet (with no subscript nor superscript) and star (with a $\star$ subscript or superscript): the mass $M$, radius $R$, rotation angle about the instantaneous shortest axis of the body $\theta$ (along with the spin rate $d\theta/dt$), compliance $\mathcal{J}$ and dynamic viscosity $\eta$. The initial values of all 14 parameters are fed into the coupled set of spin and orbital equations (which are not repeated here) C1-C10 of \cite{veretal2019b} in order to determine the motion.

The evolution equations are all secular, meaning that they compute the averaged motion over long timescales. Here, ``long'' is a quantity, such as years, which is much greater than an orbital timescale of hours or days. For the purposes of assessing detectability, computing evolution over secular timescales is desirable. Further, white dwarfs represent the endpoints of stellar evolution for the majority of stars in the Milky Way, and hence cool quiescently over $10^{10}$~yr timescales. Integrating the equations over Gyr timescales is computationally manageable because secular equations are independent of parameters, such as the mean anomaly, which pinpoint the location of the planet along the orbit.

White dwarf discs are highly unlikely to be massive enough to affect the motion of a planet at the sizes which we are considering, although the planet may perturb the disc in interesting and observationally measurable ways \citep{manetal2016,xuetal2018,manetal2019}. 

\subsection{Lorentz drift}

Now consider a magnetic white dwarf and the scenario originally envisaged by \cite{lietal1998}, \cite{wilwu2004}, and \cite{wilwu2005}. We now know that the gas required to maintain an active magnetosphere can arise from additional sources besides the interstellar medium (as explained in Section 1.1). These studies proposed that the electric currents generated by a planet travelling through a white dwarf magnetosphere will set up a unipolar inductor circuit similar to that which is seen with Jupiter and Io, where Jupiter is akin to the white dwarf and Io is akin {\rev to the planet \citep{gollyn1969}}. Then the magnetic field lines close to the white dwarf reside at the locations of radio emission zones: see Fig. 1 in each of \cite{lietal1998}, \cite{wilwu2004}, and \cite{wilwu2005}. Resulting Lorentz torques generate energy, some of which is dissipated and some of which is applied as kinetic energy to the planet. 

The magnitude of the resulting inward drift on the planet, the ``Lorentz drift'', may be estimated. A direct comparison of the drift rates from \cite{lietal1998},  \cite{wilwu2004}, \cite{wilwu2005} and \cite{broken2019} reveal that no two of their rates are equivalent, even when accounting for the different electromagnetic systems of units that were adopted\footnote{\cite{lietal1998}, \cite{wilwu2004}, and \cite{wilwu2005} all use the electrostatic ESU system, which  introduces a factor of the square of the speed of light which is not present in the SI unit system.}. 

Now we attempt to parse these differences. {\rev The results of} \cite{broken2019} differ because they consider a static dipole field and an efficiency factor based on a functional form of the current density within an asteroid from \cite{bidetal2007}. The drift rate form of \cite{lietal1998} differs from those of \cite{wilwu2004} and \cite{wilwu2005} by a numerical factor ($\pi/2$ versus 8) as well as whether to represent the magnetic field strength by the square of the stellar magnetic moment divided by the sixth power of the white dwarf radius.  \cite{wilwu2004} and \cite{wilwu2005} differ because the former includes an asynchronism term which contains a comparison of the white dwarf's angular spin speed  to the planet's orbital angular speed.

All these considerations led us to adopt the following estimate for the Lorentz drift in SI units:

\[
\left\langle
\left\langle
\frac{da}{dt}
\right\rangle
\right\rangle_{\rm Lorentz}
=
-5 \left\langle \gamma_{\star} \right\rangle B_{\star}^2 
\left( \frac{R^3}{R_{\star}} \right)
\left( \frac{R_{\star}}{a} \right)^{\frac{17}{2}}
\left( \frac{a^2}{M} \right)
\]

\begin{equation}
\ \ \ \ \ \ \ \ \ \ \ \ \ \ \ \ \ \ \ \ \ \
\times 
\left[  
1-
\left\langle
\left\langle
\frac{d\theta^{\star}}{dt}
\right\rangle
\right\rangle
\frac
{a^{3/2}}
{\sqrt{G\left(M_{\star} + M \right)}}
\right]^2
,
\label{drift}
\end{equation}

\noindent{}where the magnetic field strength is given by $B_{\star}$ and the vertically-averaged electrical conductivity in the white dwarf atmosphere is given by $\left\langle \gamma_{\star} \right\rangle$. All notations are consistent with those in \cite{veretal2019b}. 

In equation (\ref{drift}), we have (i) assumed that the drift rate is secular and slow enough to affect only the long-term change in the semimajor axis, (ii) adopted a numerical coefficient which is roughly in between those of \cite{lietal1998} and \cite{wilwu2004}, and (iii) included the asynchronism term\footnote{Despite including the asynchronism term, \cite{wilwu2004} assumed that the stellar spin period is much greater than the planetary orbital period. Now we know that white dwarf spin periods generally range from hours to days \citep{heretal2017}.}. The asynchronism term provides an additional coupling between the orbital and spin periods of motion along with those from \cite{veretal2019b}. 

In order to simulate the planet's evolution, we added equation (\ref{drift}) to Eq. C1 of \cite{veretal2019b}. However, equation (\ref{drift}) requires us to establish values for two additional and important parameters: $B_{\star}$ and $\left\langle \gamma_{\star} \right\rangle$. We devote the next two subsections to obtaining reasonable ranges for these parameters.

\subsubsection{White dwarf magnetic field strengths ($B_{\star}$)}

Observations of isolated magnetic white dwarfs show $B_{\star} = 10^3 - 10^9$ G and suggest $\dot{B}_{\star} \approx 0$ \citep{feretal2015}.  We explored this entire range of magnetic field strengths here, and further assumed $\dot{B}_{\star} \approx 0$. 
The incidence of magnetic white dwarfs is different amongst different spectral classes \citep{giaetal2012,dufetal2013,kawven2014,holetal2015,holetal2017,kawetal2019}, engulfed planets might even trigger a magnetic field in a white dwarf \citep{faretal2011,brietal2018}, and a magnetic field of at least tens of kG may explain the photometric transits and accretion rates seen in WD 1145+017 \citep{faretal2017,faretal2018b}. {\rev The highest observed magnetic fields ($\sim 10^9$ G) may result from stellar mergers \citep{brietal2015,brietal2018}}. We simply treated the value of $B_{\star}$ to be independent of other stellar parameters and the dynamical history of the system.

\subsubsection{Conductivities of white dwarf atmospheres ($\left\langle \gamma_{\star} \right\rangle$)}

{\rev Both the core of a white dwarf and the metallic core of a planet are conductive. Hence, any closed current loop connecting the magnetic white dwarf to the planet's core must generate heat in the white dwarf atmosphere. The conductivity value of the atmosphere therefore not only affects the Lorentz drift, but also energy deposition, which could lead to potentially detectable phenomena such as emission lines.}

{\rev Although most white dwarfs are of spectral type DA (hydrogen line-dominated), investigations of their atmospheric conductivity values are not widespread. Therefore, we use a sufficiently wide range of conductivities to encompass plausible values.} Table 2 of \cite{sreetal2010} provides static electric conductivity values of DB (helium line-dominated) white dwarf atmospheres for a range of equilibrium temperatures between 8000~K -- 75,000~K and plasma mass densities between $5 \times 10^{-7}$~g/cm and $5 \times 10^{-3}$~g/cm. The conductivities for all these values range over six orders of magnitude, from $\left\langle \gamma_{\ast}  \right\rangle = 6.88 \times 10^{-2}$--$9.84 \times 10^4$~Siemens/metre (or, equivalently, Mho/metre or 1/Ohm$\times$metre). However, given that the vast majority of white dwarfs are observed with effective temperatures under about 20,000 K, for our purposes we decreased this upper bound by an order of magnitude. 

Even so, the entire range of conductivities is still substantial and should be explored. Generally these conductivities decrease with increasing white dwarf cooling age; Fig.~4 of \cite{mazetal2007} indicates that this trend continues for (older) white dwarfs with effective temperatures under 8000~K. {\rev As previously mentioned}, we do not know how the conductivities of DA (hydrogen-dominated) white dwarf atmospheres compare with the results of \cite{sreetal2010}. All these considerations led us to adopt the range $\left\langle \gamma_{\ast} \right\rangle =  10^{-1}$~--~$10^4$~S/m.

\begin{figure}
\includegraphics[width=8cm]{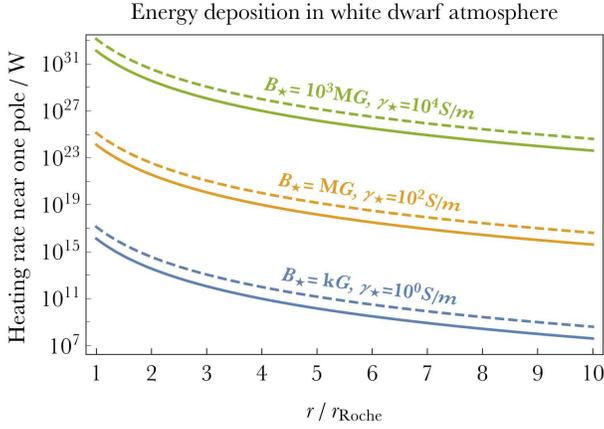}
\caption{
{\rev Energy deposited at one pole of the white dwarf due to unipolar induction with a planetary core. Solid and dashed lines respectively refer to $0.5M_{\oplus}$ and $5.0M_{\oplus}$ planets. The lower (blue) and upper (green) curves effectively bound the energy deposited by assuming limiting values of both the magnetic field strengths ($B_{\star}$) and static atmospheric conductivities ($\left\langle \gamma_{\star} \right\rangle$).
}
}
\label{energy}
\end{figure}

{\rev By using this range, we can determine how much energy is deposited in a white dwarf atmosphere. The heating rate ($H$) near one of the magnetic poles is given by Eq. 7 of \cite{lietal1998} as

\begin{equation}
H = -\frac{GM_{\star}M}{4 r^2} 
\left\langle
\left\langle
\frac{da}{dt}
\right\rangle
\right\rangle_{\rm Lorentz}
\end{equation}

\noindent{}where $r$ is the distance between the centres of the white dwarf and planetary core. 

Hence, by ignoring the asynchronism term from equation (\ref{drift}) and considering only snapshots in time, we can provide a representative range of heating rates with just a minimal set of reasonable physical parameters ($M_{\star} = 0.6M_{\star}$, $R_{\star} = 8900$ km, $\rho = 8$ g/cm$^3$, $r_{\rm Roche} = 0.68 R_{\odot}$; this choice of Roche radius is justified in Section 3.1).  Figure \ref{energy} displays the energy deposited for planetary cores of mass $M = 0.5M_{\oplus}$ (solid lines) and $M = 5.0M_{\oplus}$ (dashed lines). During planetary evolution, the heating rate will evolve with time, perhaps non-monotonically. Here, we simply plot power as a function of separation. The lower (blue) and upper (green) curves may be treated as bounds on the heat generated.
}

\section{Simulations}

{\rev Now} we describe the setup and results of our simulations. The primary challenge was selecting the most revealing regions of our large parameter space to explore. The numerical implementation involved coupling equation (\ref{drift}) to Eqs. C1-C10 of \cite{veretal2019b}. We do not repeat the details of the implementation here, except to say that we used the same generous range of the index $q$ in order to compute the eccentricity functions in their Eq. (B1). We assumed that the rheologies of both the white dwarf and planet adhere to the Maxwell model.

\subsection{Parameter choices}

The Lorentz drift is most strongly dependent on $a$, $B_{\star}$ and $\left\langle \gamma_{\ast}  \right\rangle$, the latter two parameters due to their wide ranges of plausible values. Gravitational tides between planets and white dwarfs are most strongly dependent on $a$, $M$ and $\eta$. Although all of the other parameters affect the outcomes, here we are interested only in order-of-magnitude results. Therefore, we fixed $e(0) = 0.2$, $i(0) = i'(0) = 1^{\circ}$, $M_{\star} = M_{\star}(0) = 0.6M_{\star}$, $R_{\star} = R_{\star}(0) = 8900$ km, $\theta(0) = \theta^{\star}(0) = 0^{\circ}$,  $d\theta(0)/dt = d\theta^{\star}/dt(0) = 360^{\circ}/(20 \ {\rm h})$, $\mathcal{J} = \mathcal{J}^{\star} = 0$ (1/Pa), and $\eta^{\star} = 10^7$ Pa$\cdot$s. These values are all motivated from \cite{veretal2019b}.

Regarding the physical properties of the planet, it must be a conductor in order to be detectable in radio emission. In this sense, we can better think of the planet as a planetary core. Such a core might have once hosted a gas giant, or represented most of the volume of a Super-Earth or Earth-like planet. As mentioned in Section 1.1, the scattering of large bodies like these towards a white dwarf can occur with many different potential orbital architectures \citep{debsig2002,veretal2013,voyetal2013,musetal2014,vergae2015,veretal2016,veretal2017a,steetal2018,veretal2018}. Because planetary cores are thought to be iron rich, \cite{lietal1998} assumed that $\rho = 9.4-13$ g/cm$^3$ and \cite{wilwu2005} assumed $\rho = 13.3$ g/cm$^3$ in their Fig. 2. 

However, given that the density of iron is just under $8$ g/cm$^3$, and some cores of the solar system bodies are thought to be mixtures of iron and silicates, we adopted $\rho = 8$~g/cm$^3$. Both \cite{lietal1998} and \cite{wilwu2005}, papers that were written before the revelation that super-Earths are common throughout the Galaxy\footnote{See the NASA Exoplanet Archive at exoplanetarchive.ipac.caltech.edu}, assumed the planet mass is 40 per cent of the mass of the Earth. Here we adopted two values for $M$: $M = \left\lbrace 0.5M_{\oplus}, 5.0M_{\oplus} \right\rbrace$. Consequently, the radii of our two planet cores are 4500 km and 9600 km (the latter actually being larger than the white dwarf radius). 

Unlike for the much smaller core fragment discovered orbiting a white dwarf in \cite{manetal2019}, here we did not assume that the object has any internal strength. Our planets' density, along with the mass of the white dwarf, sets the Roche radius of the white dwarf. By assuming that our planets are solid and spinning, we combined Eq. 1 from \cite{beasok2013} with Table 1 from  \cite{veretal2017b} to obtain a Roche radius (for both planets) of $r_{\rm Roche} =0.00316$~au $ = 53.2 R_{\star}$ $ = 0.68 R_{\odot}$. This location, which corresponds to an orbital period of about just two hours, is much closer than the standard Roche radius for rubble piles, at about $1.0 R_{\odot}$ and corresponding to an orbital period of about 4.5 hours \citep{vanetal2015}. The smaller core fragment from \cite{manetal2019} has an orbital period of about two hours, but is intact at its location (whereas our planets would start to break up).

The dynamic viscosity ($\eta$) of planetary cores within our solar system have been estimated to range from $10^{21} - 10^{28}$~Pa$\cdot$s \citep{huretal2018,patetal2019}. Despite this large range, the lower bound is a useful constraint, because without it, the range of consideration would increase by at least five orders of magnitude for planetary objects which are not solid cores. Here, we assumed three values for planet viscosity that span this range: $\eta = 10^{21}, 10^{24}, 10^{27}$~Pa$\cdot$s.   

For the remaining three variables, as explained earlier we adopt the ranges $B_{\star} = 10^3 - 10^9$ G and $\left\langle \gamma_{\ast} \right\rangle =  10^{-1}$~--~$10^4$~S/m, and we did so at order-of-magnitude resolution. We made no a-priori assumptions about $a$, except to sample it at multiples of $r_{\rm Roche}$. We can also consider the critical $a$ beyond which maser emission would no longer be detectable. \cite{wilwu2005} suggested that the induced potential across a detectable planet would exceed about 1 kV.\footnote{Table 2 and Figs. 3-4 of \cite{wilwu2004} suggest that much smaller induced potentials would also be detectable, but we are being conservative here.} Rewriting their equation (1) in the SI unit system then gives\footnote{Equation (1) of \cite{wuetal2002} also gives this induced potential, but with the asynchronism parameter (which we ignore here for our estimate).}

\begin{equation}
a_{\rm crit} \approx 
\left[
\frac{B_{\star} R_{\star}^3 R \sqrt{G M_{\star}}}{1 \ {\rm kV}}
\right]^{2/7}.
\label{acriteq}
\end{equation}

\noindent{}Hence, for our $0.5 M_{\oplus}$ planet, $a_{\rm crit} = 7.6 - 400 r_{\rm Roche}$ for $B_{\star} = 10^3 - 10^9$ G. For our $5.0 M_{\oplus}$ planet, $a_{\rm crit} = 9.5 - 500 r_{\rm Roche}$ for $B_{\star} = 10^3 - 10^9$ G. Fortunately, these critical semimajor axis values are high enough to not actually limit survivability; at these values of $a_{\rm crit}$ we found that the planets survive for at least 1 Gyr across the parameter space.

We adopted integration timescales of 1 Gyr because of both computational speed and due to 1 Gyr being sufficiently long compared to a typical white dwarf cooling age. Models for scattering planetary cores towards white dwarfs currently remain too unconstrained to place reliable bounds on when they arrive and how frequently they arrive, properties which would have helped us determine the ideal integration times.

\begin{figure*}
\includegraphics[width=16cm]{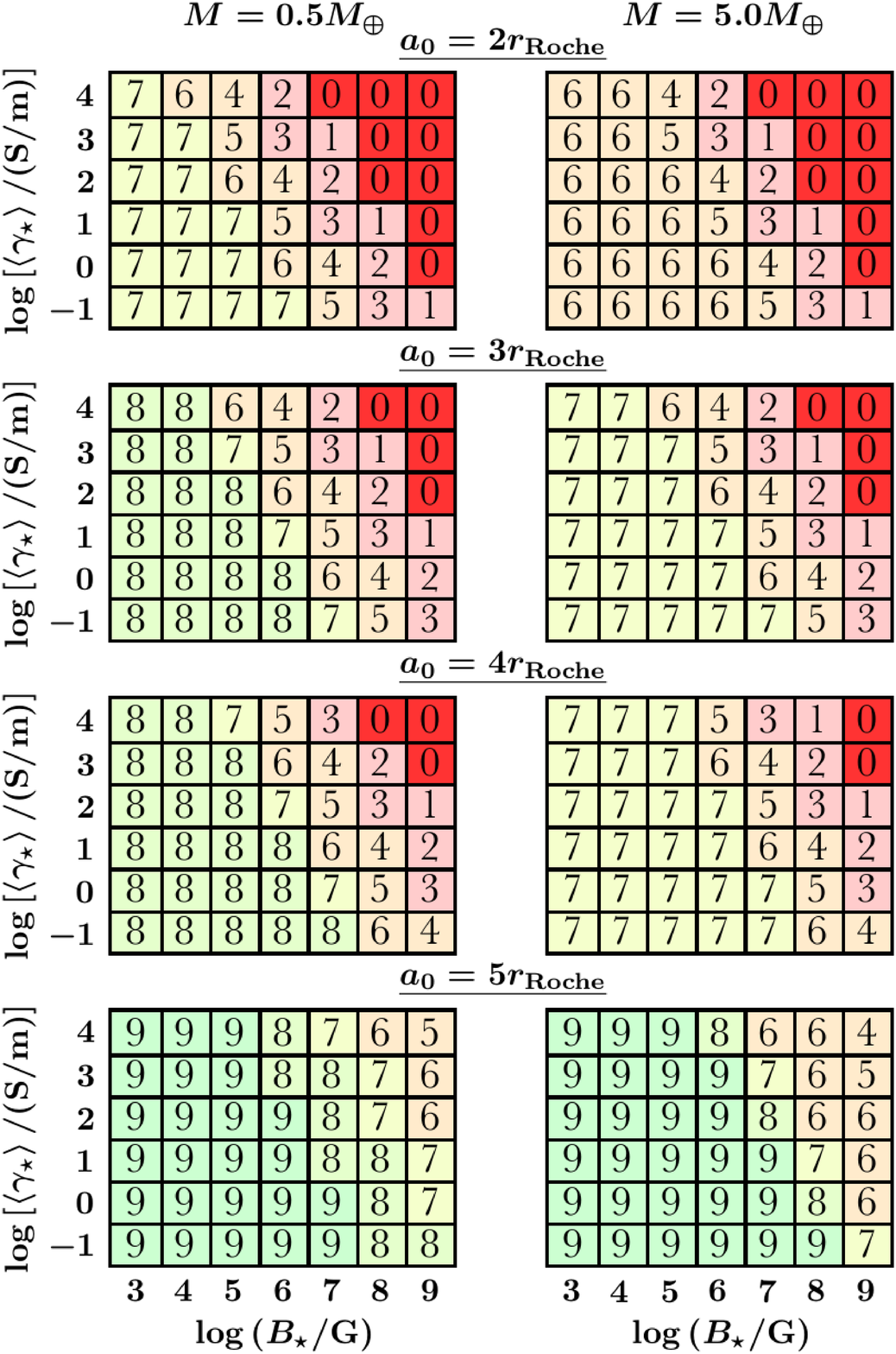}
\caption{
Survival times ($t_{\rm surv}$) of $\eta = 10^{21}$~Pa$\cdot$s planetary cores orbiting white dwarfs as a function of magnetic field strength ($x$-axis) and electrical conductivity ($y$-axis); other initial physical and orbital parameters are given in Section 2.
 A ``0''  indicates $t_{\rm surv} < 10$ yr, a ``9'' indicates $t_{\rm surv} > 10^9$ yr and an intermediate value $0 < C < 9$ indicates $10^{C} \le   t_{\rm surv} < 10^{C+1}$ yr.
}
\label{templateeta21}
\end{figure*}

\begin{figure*}
\includegraphics[width=16cm]{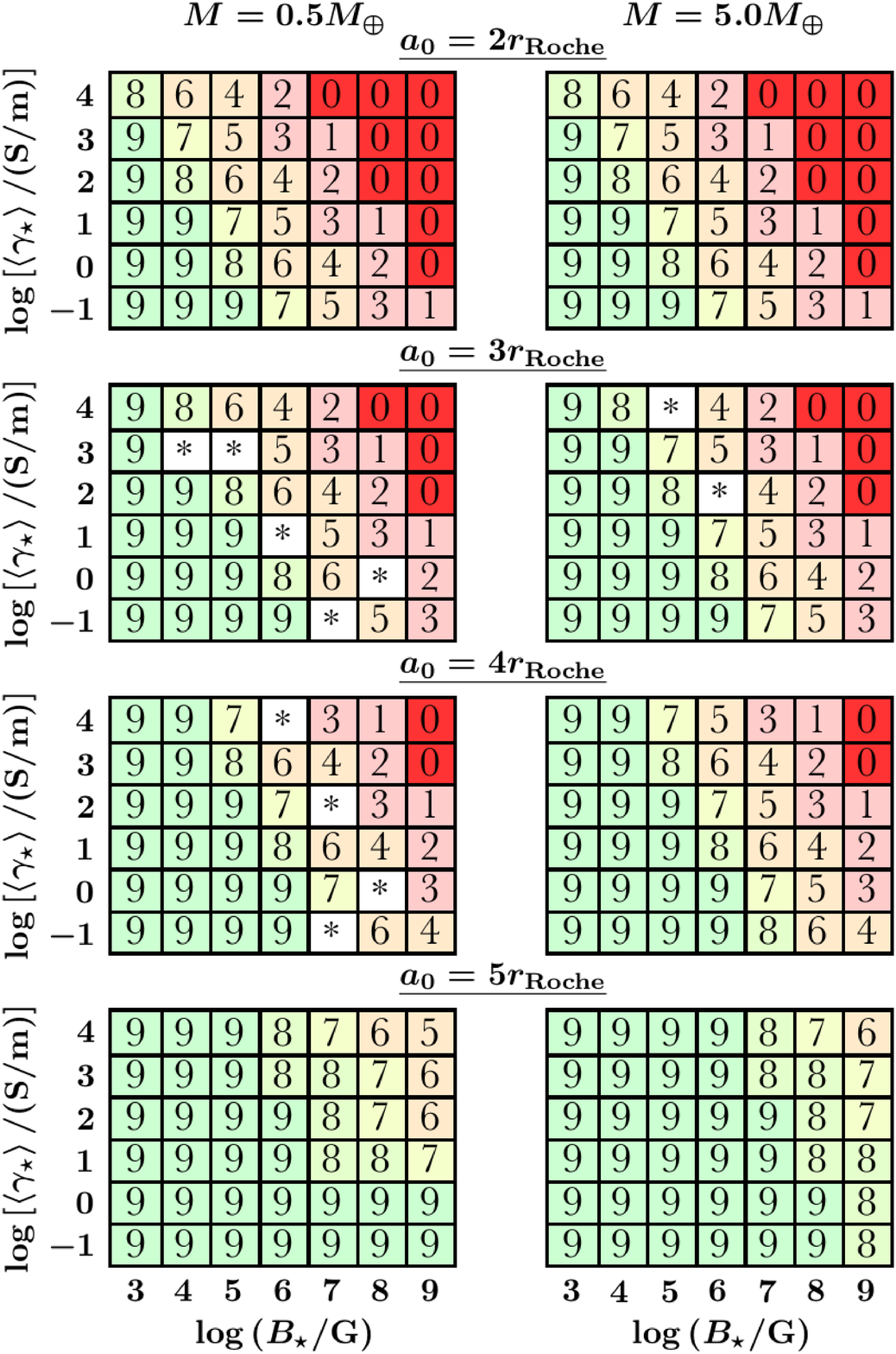}
\caption{
Like Fig. \ref{templateeta21}, except for $\eta = 10^{24}$~Pa$\cdot$s. A $\ast$ indicates that the equations become stiff for those particular initial parameters.
}
\label{templateeta24}
\end{figure*}

\begin{figure}
\includegraphics[width=8cm]{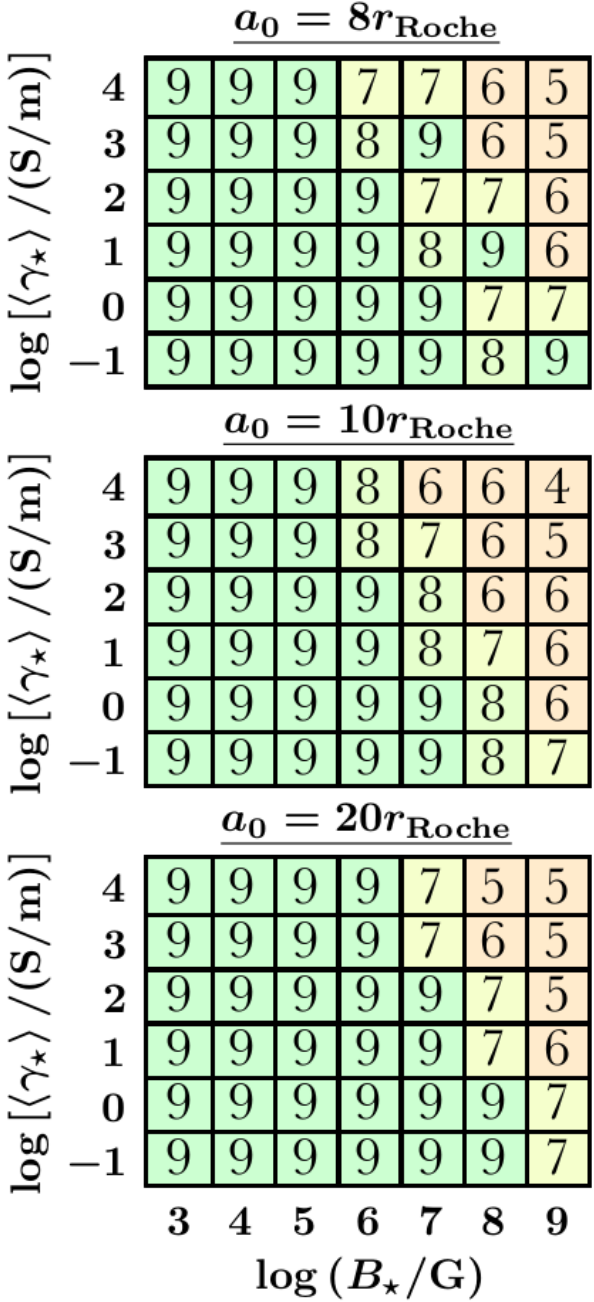}
\caption{
Continuation of the $M=5.0M_{\oplus}$ case from Fig. \ref{templateeta21}, but for higher initial $a_0$.  The non-monotonic trends are indicative of chaos from the coupled tidal and Lorentz forces.
}
\label{templateahigh}
\end{figure}

\subsection{Simulation results}

Our primarily goal is to determine approximately how long planets which are subject to both gravitational tides and the Lorentz drift can survive orbiting a white dwarf within $a_{\rm crit}$ (equation \ref{acriteq}), rather than provide details of their motion in individual cases. These details would likely change with more sophisticated models. 

Surviving planets may alter their orbits in non-obvious ways due to tides alone \citep{veretal2019b}, and especially now when coupled with an electromagnetic force. Nevertheless, conveniently we found that the planets do not move outward enough to exceed $a_{\rm crit}$. 

We assumed that a planet survives if its orbit never intersected with the sphere of radius $r_{\rm Roche}$; the planet was therefore classified as destroyed if its orbital pericentre was shifted inside of $r_{\rm Roche}$.  Because we seek just order-of-magnitude timesacle estimates, we visualize some of our results through colour-contoured number grids (Figs. \ref{templateeta21}-\ref{templateahigh}). The $x$-axis on each grid is $\log{B_{\star}}$ and the $y$-axis is $\log{\left\langle \gamma_{\ast}  \right\rangle}$. Although the product $B_{\star}^2 \left\langle \gamma_{\star} \right\rangle$ could have instead be presented on one axis as a single degree of freedom, we favour the clarity of our current presentation.  

The numbers within the grids in Figs. \ref{templateeta21}-\ref{templateahigh} correspond to survival timescales $t_{\rm surv}$ according to

\begin{itemize}

\item ``0'':  \ \ \ \ \  $t_{\rm surv} < 10$ yr.

\item ``1-8'': \ \    $10^{1-8} \le   t_{\rm surv} < 10^{2-9}$ yr.

\item ``9'':  \ \ \ \ \  $t_{\rm surv} > 10^9$ yr.

\item ``$\ast$'': \ \ \ \ \ \,  Equation set becomes stiff due to the asynchronism parameter.

\end{itemize}

\noindent{}The background colours for the numbers were chosen according to the planet's level of safety or danger. Dark (0) or light red (1-3) values indicate that destruction occurs quickly enough such that these planets are unlikely to be observed, whereas light green (8) or dark green (9) indicates that the planet may survive for at least one per cent of the white dwarf cooling age. 

Consider first the dependence of $t_{\rm surv}$ on dynamic viscosity.
In Fig. \ref{templateeta21}, we paint a picture of the $\left\lbrace B_{\star}, \left\langle \gamma_{\star} \right\rangle, M, a_0 \right\rbrace$ parameter space for the limiting case of $\eta = 10^{21}$~Pa$\cdot$s (the lowest supposed dynamic viscosity for a planetary core). This low value allows tides to dominate when the magnitudes of $B_{\star}$ and $\left\langle \gamma_{\star} \right\rangle$ are low (at the bottom lower left corner of each grid).  When $\eta$ is then increased by three orders of magnitude to $10^{24}$~Pa$\cdot$s (Fig. \ref{templateeta24}), tides no longer dominate, even for the lowest values of $B_{\star}$ and $\left\langle \gamma_{\star} \right\rangle$. A direct comparison of Figs. \ref{templateeta21} and \ref{templateeta24} demonstrates a 0-3 order-of-magnitude difference in survival timescale when the core's viscosity is increased. We also ran simulations for the $\eta = 10^{27}$~Pa$\cdot$s case, but the results are indistinguishable from those in Fig. \ref{templateeta24}.

Even when tides do not dominate the motion -- at the highest values of $B_{\star}$ and $\left\langle \gamma_{\star} \right\rangle$ (upper right corner of each grid) -- tides still play a role in the motion. Comparison of the bottom two panels within each of Figs. \ref{templateeta21} and \ref{templateeta24} reveals that the survival timescales for the highest values of $B_{\star}$ are not equivalent. Although tides do not dominate the motion here, their coupling with the Lorentz drag affects the entire dynamical system.

The next dependence to note in Figs. \ref{templateeta21} and \ref{templateeta24} is on $a_0$. Both tides and the Lorentz force are steep functions of distance to the white dwarf.  For low-viscosity planets (Fig. \ref{templateeta21}), we find that none of them survive for at least 1 Gyr unless $a_0 \gtrsim 5r_{\rm Roche}$. The survival times of these planets at the highest values of $B_{\star}$ and $\left\langle \gamma_{\star} \right\rangle$ increase by at least four orders-of-magnitude just by incrementing $a_0$ from $4r_{\rm Roche}$ to $5r_{\rm Roche}$. A similar trend is seen for more viscous cores (Fig. \ref{templateeta24}). In general, planets survive for longer the further away they start out from the star. Because a distance of $5r_{\rm Roche}$ is under $a_{\rm crit}$ for all values of $B_{\star}$, this interval ($5r_{\rm Roche}, a_{\rm crit}$) contains potentially detectable planets, and for over a large region of the remaining parameter space. 

Although $t_{\rm surv}$ appears to increase monotonically with $a_0$,  Fig. \ref{templateahigh} illustrates that this trend is only secular. In particular, the upper panel of the figure exhibits a few cases where increasing $B_{\star}$ actually increases the survival time, and how $t_{\rm surv}$ actually decreases from the top to the middle panel for the most magnetically active system. This behaviour is due entirely to the coupling of gravitational tides with the Lorentz drift: we re-ran the same $a_0 = 8r_{\rm Roche}$ grid simulations without tides, and the result was a grid entirely filled with ``9''s. In other words, even for the most magnetically active systems, the Lorentz drift on its own would not drag the planet into the white dwarf within 1 Gyr for $a_0 \ge 8r_{\rm Roche}$.

The final correlation to consider is with planet mass, which strongly affects gravitational tides by themselves. Here, however, comparison of the right and left columns of each of Figs. \ref{templateeta21} and \ref{templateeta24} show differences on the order of just one order-of-magnitude for $t_{\rm surv}$ as the planet mass is varied by one order-of-magnitude. 

\section{Applicability to other planetary systems}

The set of equations we used to obtain our results may also be applied to other planetary systems where a unipolar inductor circuit has been established between a star and a solid planet that can both be characterised with Maxwell rheologies. Our model can thus help characterise planetary systems which do not fall under the remit of the few other investigations which have so far considered the interplay between gravitational tides and magnetic torques \citep{bouceb2015,stretal2017}. These studies focused on protostars and hot Jupiters, T-Tauri stars and hot Jupiters, K stars and hot Jupiters, and M-dwarfs and Earths.  Hot Jupiters require a fundamentally different treatment of gravitational tides than do Earths \citep{ogilvie2014,mathis2018}. Further, even amongst rocky bodies, there is a wide variety of potential star-planet tidal formulations; the one we have adopted from \cite{veretal2019b} contains an arbitrary frequency dependence on the quality functions, thereby avoiding pitfalls highlighted by \cite{efrmak2013} and \cite{makarov2015}.

Although our numerical implementation would not change when applied to a different type of host star, the physical parameters of the star would. One of the most consequential changes would be for $\eta^{\star}$, whose value for white dwarfs ($10^7$ Pa$\cdot$s) is many orders of magnitude lower than what we might expect for other types of stars. Hence, in other systems, both stellar and planetary tidal forces may be comparable, complicating the trends observed in this study when coupled with the Lorentz drift.

\section{Applicability to GD 356 and GD 394}

{\rev Although $\eta^{\star}$ could change by many orders of magnitude for other types of stars, we do not expect it to vary significantly amongst different white dwarfs. Two white dwarfs which represent tantalising candidates for potentially hosting planetary cores are GD 356 \citep{feretal1997,wicetal2010} and GD 394 \citep{dupetal2000,wiletal2019}. GD 356 is known to be magnetic, whereas a magnetic field has not yet been detected in GD 394.

The energy source which heats GD 356 remains unknown. However, \cite{wicetal2010} strengthened the unipolar circuit hypothesis by using {\it Spitzer} observations to robustly constrain the mass of a planetary companion to be under 12 Jupiter masses. Also, that star harbours a magnetic field strength on the order of 10 MG. Based on our simulations, this (high) value places further constraints on a potential companion: it must reside beyond about $3-4 r_{\rm Roche}$ in order to survive for longer than about $10^{7-8}$ yr. Further, the spin period of the white dwarf is about two hours, an order of magnitude shorter than the period we adopted here. Consequently, the region of parameter space where the tidal and Lorentz contributions are comparable would differ from the region identified in the simulations performed here, but most likely lie within $3-4 r_{\rm Roche}$ and hence negligibly affect a detectable planet.

For GD 394, flux variations in the EUV led \cite{dupetal2000} to propose that the white dwarf is accreting locally, in a surface accretion spot. More recent observations \citep{wiletal2019} show no evidence for this spot, perhaps indicating that it is no longer currently active. Regardless, heating from the unipolar circuit does occur locally, at the ``feet'' of the flux tube. Therefore, searching for localized heating in other white dwarfs is a potential pathway for discovering magnetic planetary cores.
}

\section{Summary}

White dwarf planetary systems are ubiquitous, with an occurrence rate similar to that of main sequence planetary systems. Although observations of these evolved systems are dominated by planetary remnants in the form of metal pollution and circumstellar dust and gas, individual objects have now been seen orbiting these types of stars \citep{vanetal2015,manetal2019}. The most recent find is a high-density metallic core fragment orbiting within one Solar radius of the white dwarf.

Prospects for finding additional planets have received a major boost with an order of magnitude increase in the known population of white dwarfs in the year 2018 alone from {\it Gaia} Data Release \#2 \citep{genetal2019}, as well as the successful launch of {\it TESS} and commissioning of {\it  LSST} \citep{corkip2018,lunetal2018}. Further, the spectroscopic technique utilized by \cite{manetal2019} to discover the core fragment may be applied to other white dwarfs which contain gaseous discs. This disc population is also expected to increase, particularly with the expected launch of {\it Euclid}.

As the field of post-main-sequence planetary system science grows, so does the need for models which can assess the likelihood of discovering planets at particular locations and with particular physical properties. In this paper, we have revisited the possibility of a radio-loud conducting planetary core orbiting a white dwarf through a closed unipolar circuit which survives long enough to be detected \citep{lietal1998,wilwu2004,wilwu2005}. We provided a major update to this model by (i) incorporating gravitational tides from \cite{veretal2019b} into the computation in combination with Lorentz drift, and (ii) by performing simulations across the entire range of observable magnetic field strengths and expected electrical conductivities of white dwarf atmospheres. We found:

\begin{itemize}

\item Planetary survival times of $t_{\rm surv} > 10^8$ yr are common in regions of parameter space where the induced potential across the planet exceeds 1 kV.

\item Lorentz drift (equation \ref{drift}) and gravitational tides (Eqs. C1-C10 of \citealt*{veretal2019b}) are coupled. Although each force dominates in a different region of parameter space, in other regions they provide feedback on each other. 

\item Planet migration may be inwards or outwards, but nearly always is inwards.

\item $t_{\rm surv}$ most strongly depends on the planet's dynamic viscosity $\eta$, the star's magnetic field strength $B_{\star}$ and the initial star-planet distance $a_0$. The next strongest dependencies are on the white dwarf's electrical conductivity $\left\langle \gamma_{\star} \right\rangle$ and the planet mass $M$.

\item $t_{\rm surv}$ also is dependent on, but usually weakly, the spins of both the white dwarf and planet, the planet's density, the star's mass and radius (which are well constrained for white dwarfs; \citealt*{treetal2016}), the star's dynamic viscosity (as long as it is not too many orders-of-magnitude higher than $10^7$~Pa$\cdot$s), both the planet's and the star's compliances (as long as they do not vary too far from zero), and the eccentricity and inclination of the initial orbit.

\item Planetary tides dominate the motion when $B_{\star} \approx$ kG-MG, $\eta \sim 10^{21}$~Pa$\cdot$s and $a_0 \le 4r_{\rm Roche}$.

\item Planetary tides never dominate the motion when $\eta \gtrsim 10^{24}$~Pa$\cdot$s.

\item The highest observed magnetic fields ($\sim 10^3$ MG) rarely allow for $t_{\rm surv}$ to reach as high as $10^8$ yr, even at distances of tens of $r_{\rm Roche}$.

\end{itemize}

\section*{Acknowledgements}

{\rev We thank the referee for helpful comments which have improved the manuscript.}
This research was supported in part by the National Science Foundation under Grant No. NSF PHY-1748958 through the Kavli Institute for Theoretical Physics programme ``Better Stars, Better Planets''. DV gratefully acknowledges the support of the STFC via an Ernest Rutherford Fellowship (grant ST/P003850/1). The Centre for Exoplanets and Habitability is supported by the University of Warwick, and the Center for Exoplanets and Habitable Worlds is supported by the Pennsylvania State University and the Eberly College of Science.

\label{lastpage}
\end{document}